# A Silicon Photonic Integrated Frequency-Tunable Microwave Photonic Bandpass Filter

Weifeng Zhang and Jianping Yao
Microwave Photonics Research Laboratory
School of Electrical Engineering and Computer Science, University of Ottawa
Ottawa, ON K1N 6N5, Canada
jpyao@eecs.uottawa.ca

***Abstract*—We report a fully integrated microwave photonic passband filter on a silicon photonic chip. The on-chip integrated microwave photonic filter (IMPF) consists of a high-speed phase modulator (PM), a thermally tunable high-Q micro-disk resonator (MDR), and a high-speed photodetector (PD). The bandpass filtering function of the IMPF is implemented based on phase modulation and phase-modulation to intensity-modulation conversion, to translate the spectral response of an optical filter, which is the MDR in the IMPF, to the spectral response of a microwave filter. By injecting a CW light from a laser diode (LD) to the PM, a passband IMPF with a broad tuning range is demonstrated. Thanks to the ultra-narrow notch of the MDR, a passband IMPF with a 3-dB bandwidth of 2.3 GHz is realized. By thermally tuning the MDR, the center frequency of the IMPF is tuned from 7 to 25 GHz, achieving a tuning range of 18 GHz at a power consumption of 1.58 mW. This successful implementation of an IMPF marks a significant step forward in full integration of microwave photonic systems on a single chip and opens up avenues toward real applications of microwave photonic filters.**

## I. INTRODUCTION

Due to the low loss and large bandwidth offered by modern photonics, microwave signal processing in the optical domain has been a topic of interest in the last 20 years [1, 2]. As one of the basic signal processing blocks, a microwave photonic filter (MPF) has been extensively researched and numerous solutions have been proposed [3, 4]. Most of the MPFs proposed in the past are implemented based on discrete fiber-based optical components, making the filters very bulky, expensive and high-power consuming. For many applications, however, it is highly desirable that an MPF is implemented using photonic integrated circuits (PICs) [5]. Recently, a tunable MPF that is monolithically integrated on an indium phosphide substrate has been demonstrated [6]. In its implementation, the MPF includes all the necessary elements including a laser source, a modulator, an optical filter, and a photodetector (PD). The tuning of the filter is achieved by controlling electrical currents into the micro-heaters to tune the optical filter. The main problems associated with this MPF implementation are its large size due to the low index contrast in a III-V platform and high fabrication cost due to the multiple regrowth steps in the III-V fabrication process [7]. On the contrary, silicon photonics exhibits overwhelming advantages in term of the compact footprint and low-cost fabrication with high volume production [8]. Recently, extensive efforts have been directed to the use of silicon photonic technology in the design and implementation of MPFs [9, 10]. However, to date, a fully integrated MPF on a silicon chip has not been reported.

In this paper, we report a fully integrated microwave photonic passband filter on a silicon photonic chip. The bandpass filtering function of the integrated microwave photonic filter (IMPF) is designed based on phase-modulation to intensity-modulation (PM-IM) conversion, which translates the spectral response of an optical filter to the bandpass of the IMPF. To implement such an IMPF, a high-speed phase modulator (PM), a thermally tunable high-Q micro-disk resonator (MDR), serving as the optical filter, and a high-speed PD are designed and integrated on a single silicon photonic chip. By using the silicon photonic fabrication technology, the IMPF is manufactured. By injecting a CW light from a laser diode (LD) into the chip, a passband IMPF with a broad frequency-tunable range is demonstrated experimentally. Thanks to the ultra-narrow notch of the MDR, a passband IMPF with a 3-dB bandwidth of 2.3 GHz is realized. By thermally tuning the MDR, the center frequency of the IMPF is continuously tuned from 7 to 25 GHz, achieving a tuning range of 18 GHz at a power consumption of 1.58 mW.

The proposed IMPF realizes an on-chip integration of all components except the LD, which largely reduces the system cost and significantly increase the system stability. Thanks to the high Q-factor and strong tuning capacity of the MDR, the proposed IMPF has a broad tuning range and an ultra-low power consumption. The successful demonstration of this silicon photonic IMPF represents a significant step forward in fully integration of microwave photonic subsystems and systems on a single silicon photonic chip and opens up avenues toward real applications of MPFs.

## II. INTEGRTAED MICROWAVE PHOTONIC FILTER

Fig. 1 illustrates the perspective view of the proposed IMPF, which consists of a high-speed PM, a thermally tunable high-Q factor MDR, and a high-speed PD. A CW light from an LD is coupled into the chip via an input grating coupler. In order to minimize the chip footprint and reduce the bending loss, single-mode wire waveguide structure is mostly used to route the optical signal in the chip. The input CW light is firstly guided to the high-speed PM, where the light is phase-modulated to generate a phase-modulated optical signal at the output of the PM. Then, the phase-modulated optical signal is sent to the high-Q MDR. A Y-branch coupler is connected to the output of the MDR, to split the optical signal equally into two channels. One channel of the optical signal is guided to the high-speed PD, where optical to electrical conversion is performed, to generate a microwave signal, and the other channel is routed to an output grating coupler to couple the light out of the chip, to monitor the optical spectrum. Since a fiber array is used for input and output coupling, the input and output grating couplers are deployed with a center-to-center spacing of 127 μm, which is equal to the physical spacing between two adjacent fibers in the fiber array. It is known that PM-IM conversion using an optical filter to filter out one sideband of a phase-modulated optical signal would translate the spectral response of the optical filter to the spectral response of the microwave filter [4]. In our proposed IMPF, if one of the two sidebands of the phase-modulated optical signal falls in the notch of the MDR, the phase-modulated signal is converted to an intensity-modulated signal, and at the PD the microwave signal is recovered. On the other hand, if both sidebands are applied to the PD, due to the out-of-phase nature of the microwave signal, the beating between the upper sideband and the optical carrier will fully cancel the beating between the lower sideband and the optical carrier, thus no microwave signal will be generated. The entire operation corresponds to a microwave bandpass filter with the spectral response of the microwave bandpass filter being translated from the spectral response of the optical filter. The configuration of our proposed IMPF is very simple, which is advantageous to ensure the IMPF to have a small optical insertion loss and to have a small footprint.

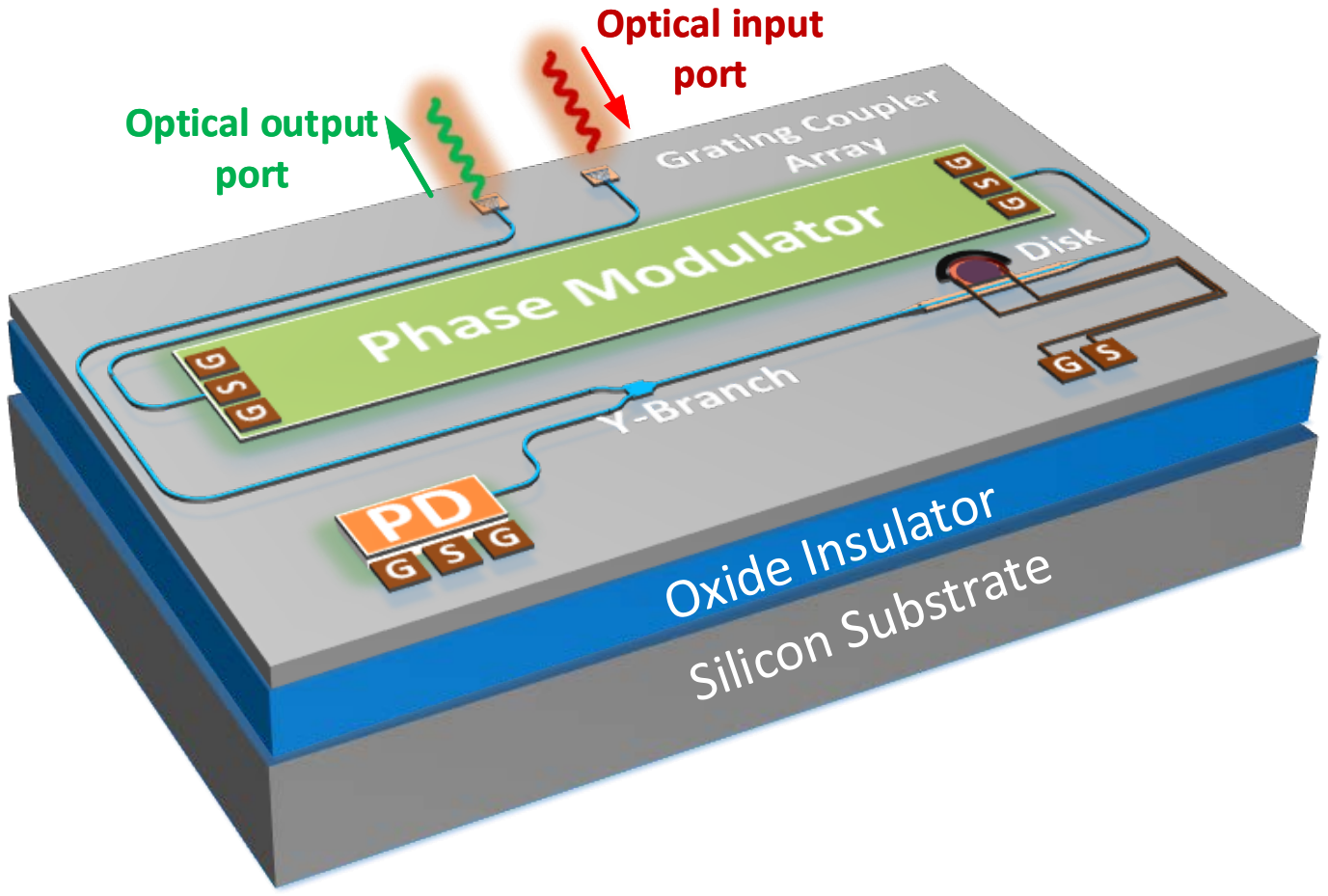

Fig. 1. Perspective view of the proposed IMPF on a silicon photonic chip

The proposed IMPF is fabricated using a CMOS compatible process with 248-nm deep ultraviolet lithography. Fig. 2(a) shows an image of the fabricated IMPF captured by a microscope camera, and Fig. 2(b) shows the layout of the IMPF. Fig. 2(c) shows an image of the input and output grating coupler, Fig. 2(d) shows an image of the high-speed PD, Fig. 2(e) shows an image of the Y-branch coupler, and Fig. 2(f) shows an image of the thermally tunable high-Q MDR.

The high-speed PM is used to modulate the incoming microwave signal on the CW light from the LD. In our proposed IMPF, the PM is designed to have a traveling wave structure where a lateral pn junction is incorporated in the rib waveguide to achieve high-speed modulation. To ensure a good match of the microwave and optical velocities, coplanar metal strips with a GSG configuration are utilized for the microwave signal. The total length of the PM is 4.5 mm. Fig. 3 shows the measured optical spectrum of a modulated optical signal at the output grating coupler using the on-chip high-speed PM with the microwave signal applied to the PM via a pair of high-speed microwave probes. The measured optical spectrums confirm that the microwave signal is modulated on the optical carrier. To monitor the sidebands of the phase-modulated optical signal in the PM performance evaluation, the wavelength of the optical carrier is chosen to be far away from the resonance wavelength of the MDR. Fig. 3(a), (b) and (c) shows the measured optical spectrum of the modulated optical signal when the frequency of the incoming microwave signal is at 7, 16 and 25 GHz, respectively. As can be seen, the power difference between the optical carrier and first-order sideband is increased from 11.9 to 12.7, and to 18.6 dBm. As the frequency of the microwave signal is increasing, the power of the sidebands is decreased, which illustrates that the modulation index at a high frequency is becoming smaller.

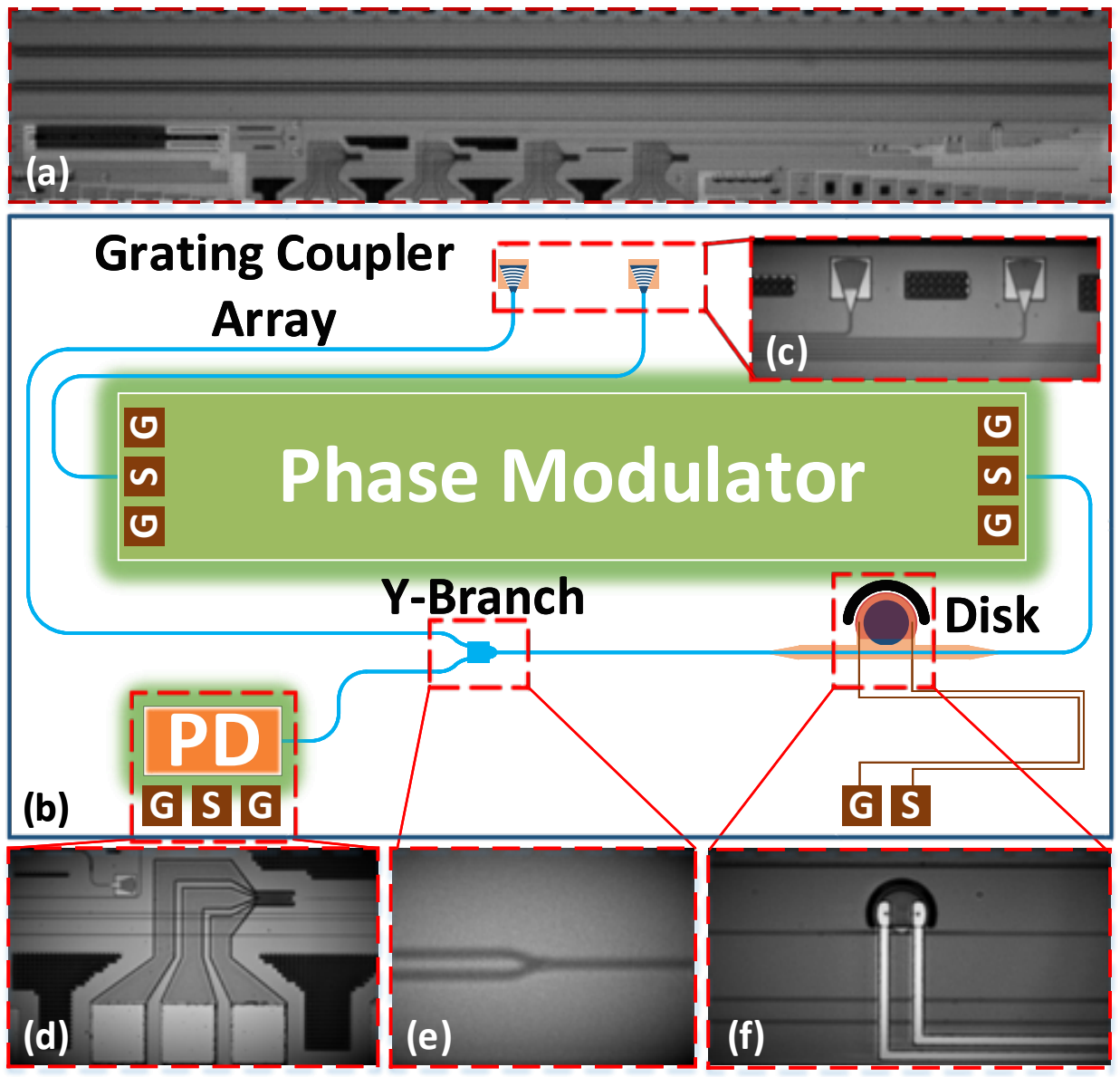

Fig. 2. (a) Image of the fabricated IMPF captured by a microscope camera, (b) schematic layout of the designed IMPF; image of (c) the grating coupler array, (d) the high speed PD, (e) the Y-branch coupler, and (f) the tunable MDR.

The thermally tunable MDR is used to act as a tunable optical filter, which is the key component to determine the spectral response of the microwave photonic filter. In our proposed IMPF, a specifically-designed MDR is used. The MDR has an all-pass configuration with a radius of 10 μm, and an additional slab waveguide is employed to wrap the disk and the lateral sides of the bus waveguide, which is able to weaken the impact of the disk sidewall roughness on the confined optical field and to strengthen the optical coupling between the disk and the bus waveguides. To ease the design, the widths of the slab waveguide around the disk and the bus waveguide are kept identical of 200 nm, and in the coupling region the two slab waveguides fully overlap. Fig. 4(a) shows the measured transmission spectrum of the fabricated MDR using an optical vector analyzer (LUNA OVA CTe) and the inset gives a zoom-in view of the notch at a wavelength of 1532.20 nm. As can be seen, the fundamental whispering-gallery-mode has a free spectral range (FSR) of 10.6 nm and the notch at the wavelength of 1532.20 nm has a 3-dB bandwidth of 18 pm, corresponding to a Q-factor of around $8.5\times10^4$, and an extinction ratio as high as 22 dB. To tune the MDR, a high-resistivity metallic micro-heater is placed on top of the disk. Fig. 4(b) shows the measured U-I curve of the micro-heater. The resistance of the micro-heater is estimated to be 30.2 Ω. By applying a DC current to the micro-heater, due to the thermal-optic effect, the refractive index of the silicon is increased, and thus the resonance wavelength of the MDR is red-shifted. Fig. 4(c) shows the measured transmission spectrum of the MDR when a different current is applied to the micro-heater. As can be seen, as the current increases, the notch of the MDR is red-shifted. When the applied electrical power to the micro-heater is 132.5 mW, the red-shift amount of the notch is measured to be 9.2 nm. In addition, during the red-shift, the extinction ratio of the notch is changed, which is caused by the waveguide-dispersion-induced coupling condition change for a different optical wavelength. The key advantage of using the MDR is its high Q-factor and its strong thermal tunability with an ultra-low power consumption.

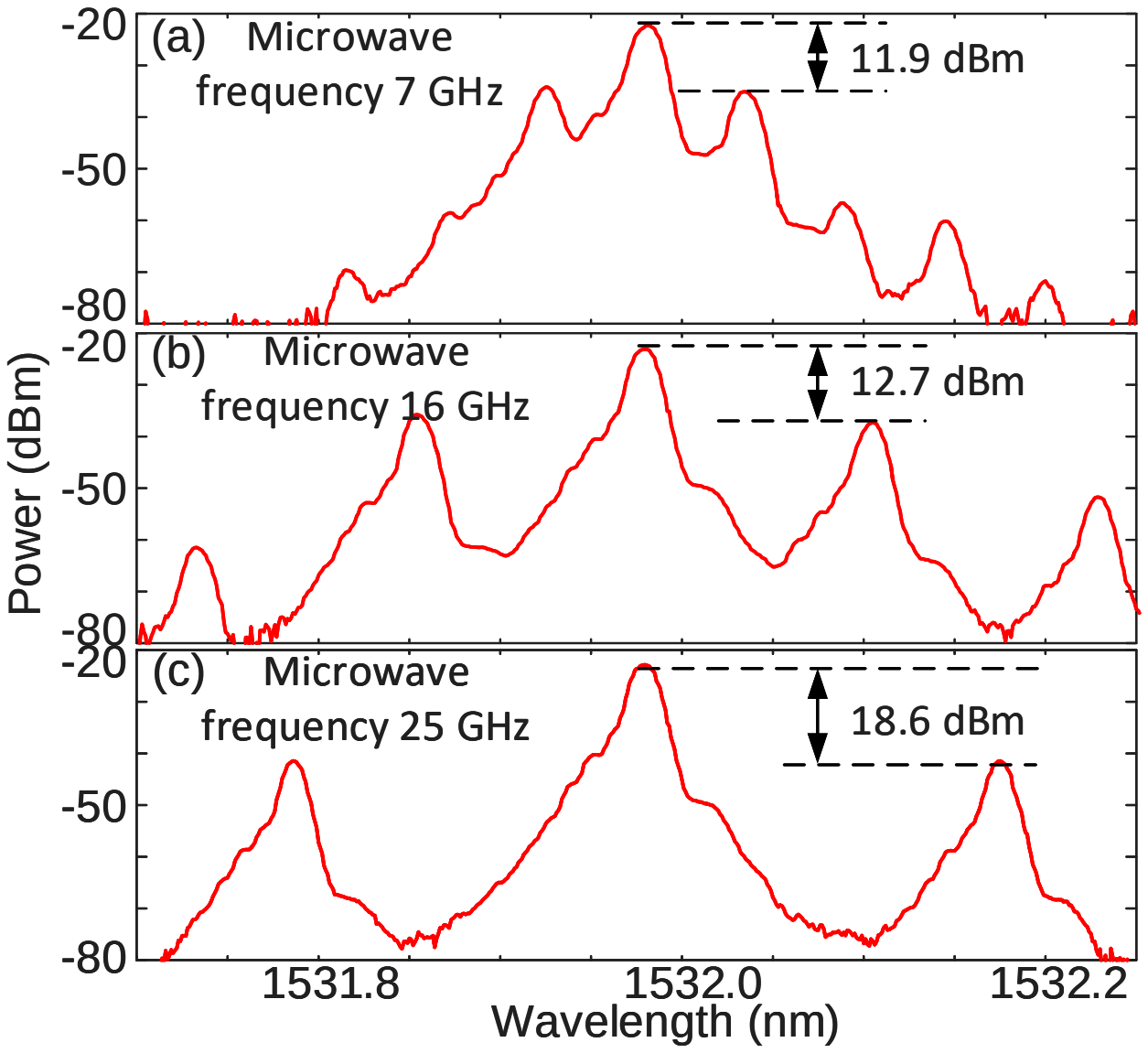

Fig. 3 Optical spectrums of the modulated optical signal using the on-chip high-speed PM when the microwave signal applied to the PM has a frequency of (a) 7, (b) 16 and (c) 25 GHz.

The high-speed PD is used to convert the modulated optical signal to a microwave signal. In our proposed IMPF, a Germanium-on-silicon PD is employed, which has a vertical Germanium PIN junction. The PD has a footprint of 56 µm in length and 24 µm in width, and a GSG configuration is used to collect the microwave signal. Fig. 5 shows the measured current at the output of the PD when a different input optical power is applied. As can be seen, the PD has a good linearity between the output current and input optical power for a given responsivity at a given reverse-bias voltage. In addition, as the reverse bias voltage increases, the responsivity is also increased until the junction breaks down. For the PD on the chip, the average on-chip responsivity is measured to be 6.78 mA/W and the dark current is 0.581 µA, which are measured at a reverse-bias voltage of 1 V.

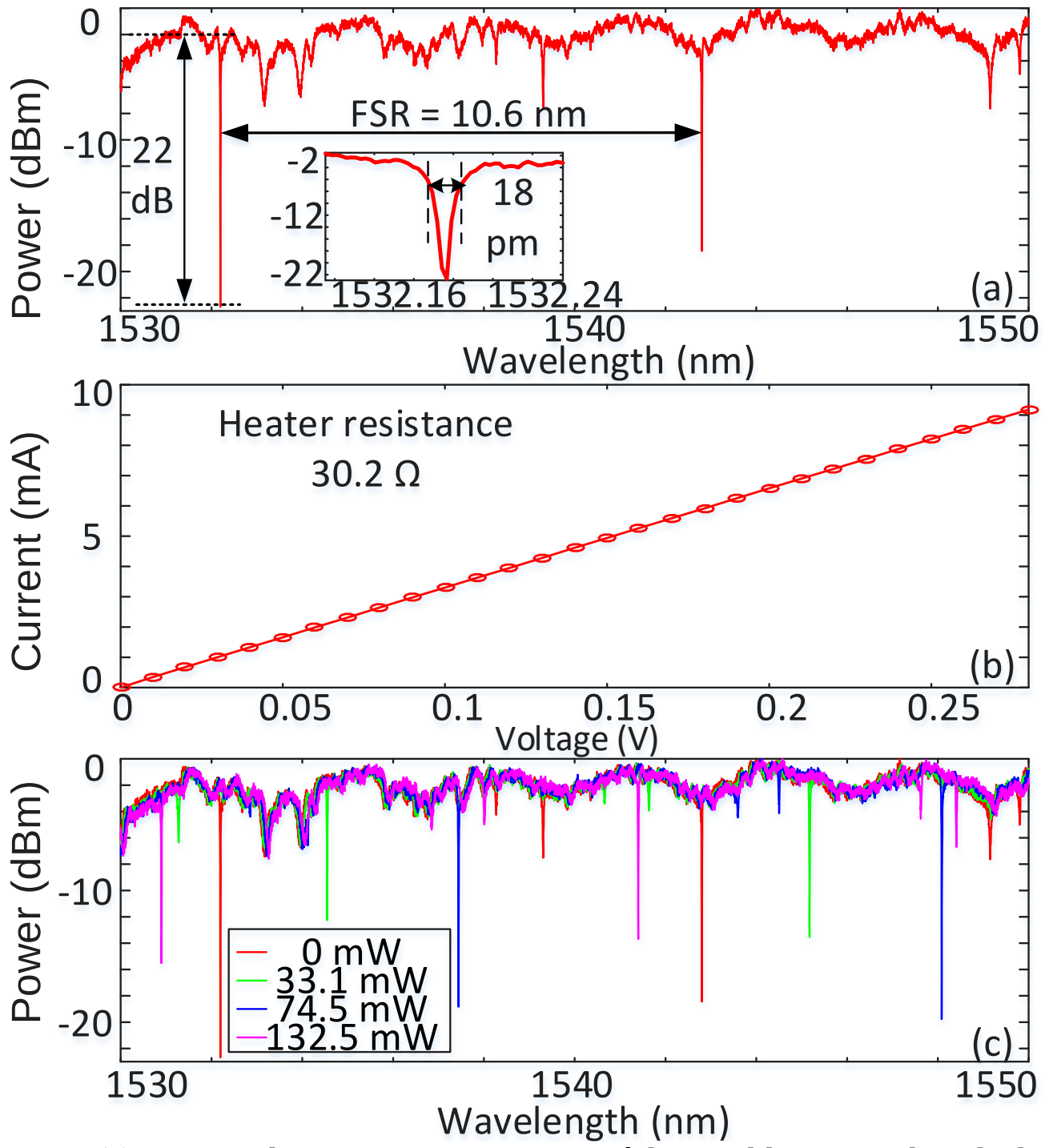

Fig. 4. (a) Measured transmission spectrum of the tunable MDR when the bias voltage is zero; (b) U-I curve of the micro-heater; and (c) thermal tunability of the MDR when the micro-heater is injected with different currents.

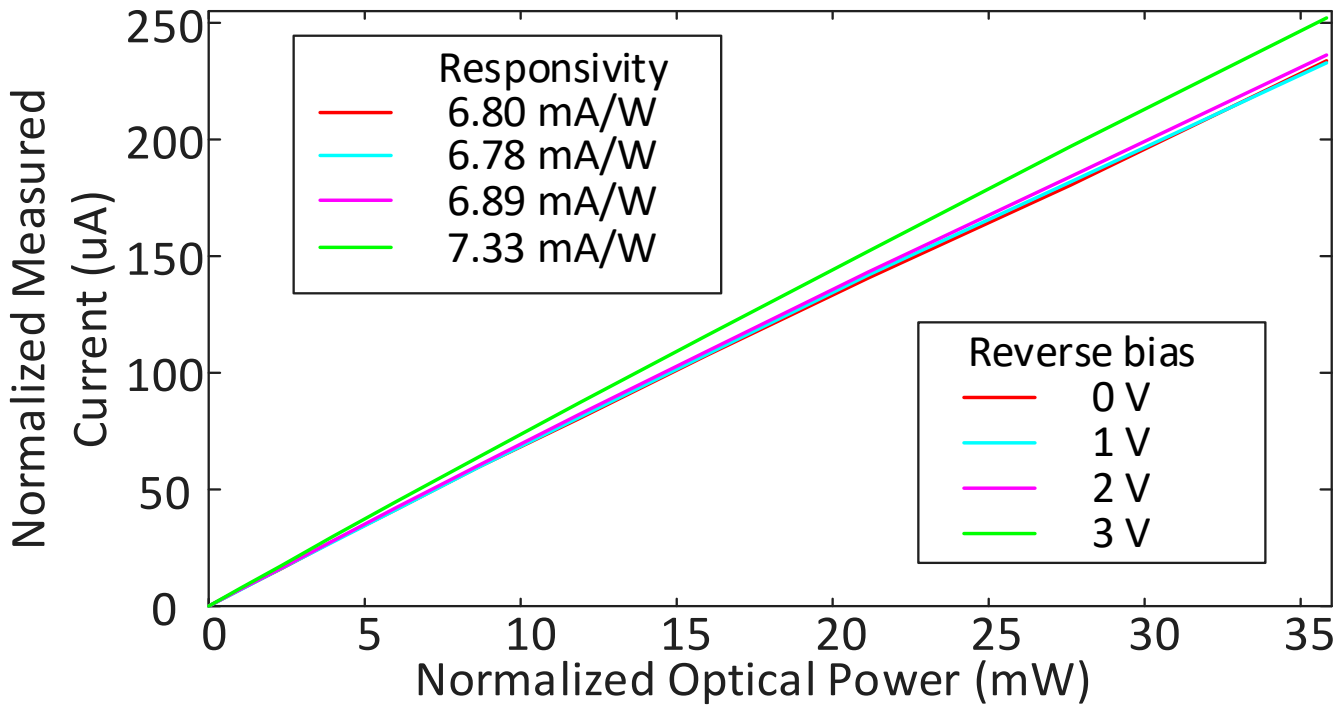

Fig. 5. Measured current of the high-speed PD when the input optical signal has a different optical power.

## III. Experimental Setup And Results

An experiment is performed to demonstrate a tunable bandpass MPF using the fabricated chip. Thanks to the high-level integration of the chip, the experimental set-up is very simple. Only an additional LD and a polarization controller (PC) are needed. The LD is used to generate a CW light with a wavelength of 1531.40 nm and a power of 7 dBm, and the PC is used to adjust the state of polarization (SOP) of the input light to the PM to minimize the polarization-dependent loss. In addition, a pair of microwave probes with a GSG configuration is used to apply an incoming microwave signal to the PM, a microwave probe with a GSG configuration in combination with a bias tee is used to apply a reverse bias voltage to the PD and to collect the recovered microwave signal from the PD, and a DC bias probe is used to apply a DC current to the micro-heater of the MDR. Note, in the experiment, due to the very close proximity of the probes, the mutual coupling between the signals from the microwave probes used to apply the incoming microwave signal to the PM and to collect the recovered microwave signal at the output of the PD happens and strong electromagnetic interference (EMI) is generated, which heavily impacts the filter performance. To overcome the EMI problem, in the experiment an off-chip commercial PD is used instead of the on-chip PD. Although an off-chip PD is used in the experiment, the on-chip PD is able to play its role after high-speed packaging to reduce the EMI. Fig. 6 is a picture showing the experimental setup captured by a camera. A fiber array is used to couple the CW light from the LD into the chip and to couple the optical signal from the MDR out of the chip. A pair of microwave probes with a GSG configuration is used for applying the incoming microwave signal to the PM, and a DC bias probe is used to apply a DC current to the micro-heater for the tuning of the MDR.

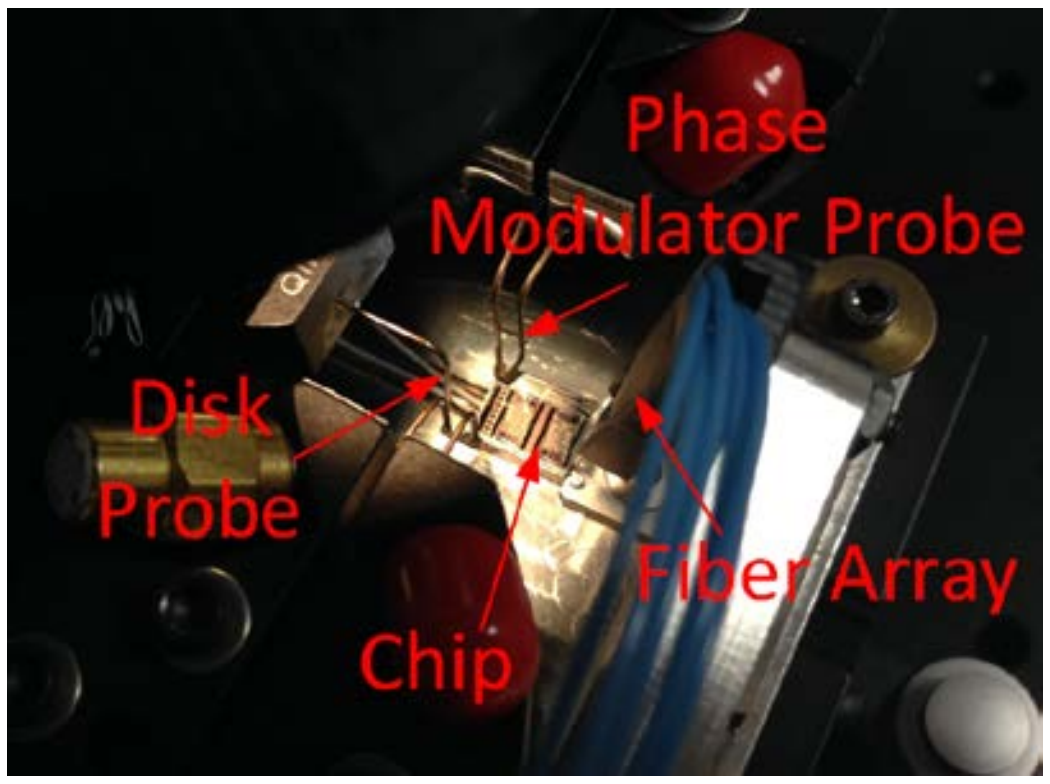

Fig. 6 Image of the experimental set-up captured by a camera.

Fig. 7(a) shows the measured frequency response of the IMPF with a center frequency of 13 GHz when a DC current of 2.99 mA (corresponding a power of 0.27 mW) to the the micro-heater. This IMPF is measured to have a 3-dB bandwidth of 2.3 GHz, which matches well with the 3-dB bandwidth of the notch of the MDR, and an extinction ratio of 17 dB. Fig. 7(b) shows the measured optical spectrum of the modulated optical signal when the incoming microwave signal frequency is 13 GHz. As can be seen, the power difference between the first-order sideband is as large as 27 dB, which is caused by locating the upper sideband at the notch of the MDR. Thus, the upper sideband is heavily suppressed, and PM-IM conversion is implemented. By thermally tuning the MDR, the center wavelength of the notch is shifted, which leads to the tuning of the IMPF. Fig. 7(c) shows the measured frequency responses of the IMPF with the center frequency tuned from 7 to 25 GHz when a different DC current is applied to the micro-heater. As can be seen, a tuning range as broad as 18 GHz is achieved with a power consumption as small as 1.58 mW, which confirms the key advantage of the IMPF in the terms of a broad frequency-tunable range and low power consumption.

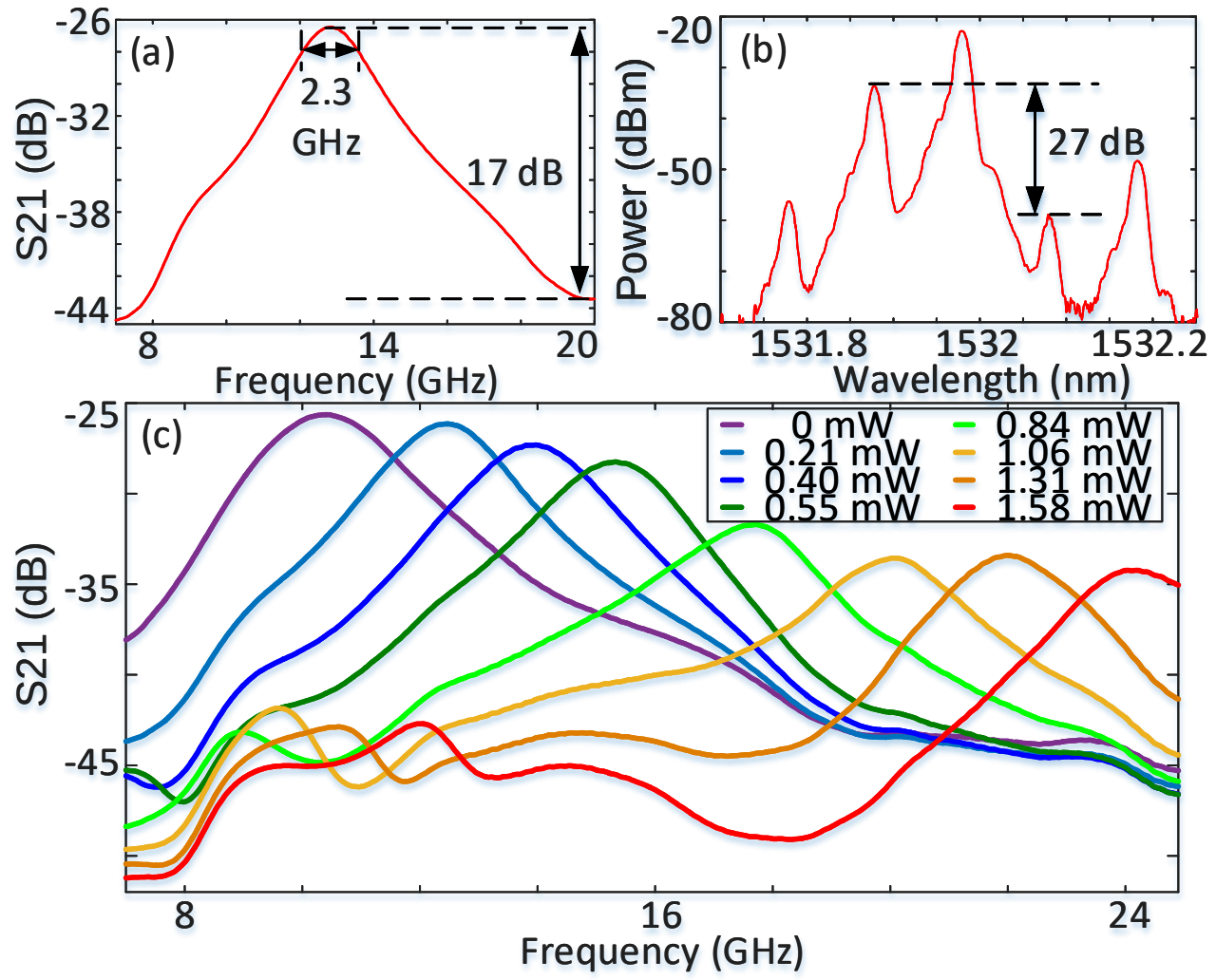

Fig. 7 Experimental result: (a) frequency response of the filter with a center frequency of 13 GHz; (b) optical spectrum of the modulated optical signal when the microwave signal frequency is 13 GHz; (c) measured frequency responses of the filter with the center frequency tuned from 7 to 25 GHz.

## IV. SUMMARY

A fully integrated passband MPF was designed, fabricated and demonstrated on a silicon photonic chip. Thanks to the ultra-narrow notch of the MDR, a passband IMPF with a 3-dB bandwidth of 2.3 GHz was realized. By thermally tuning the MDR, the center frequency of the IMPF was tuned from 7 to 25 GHz, achieving a tuning range of 18 GHz at a power consumption as low as 1.58 mW.